\documentclass[letter,longauth]{aa}

\usepackage[varg]{txfonts}

\usepackage{graphicx}
\usepackage{epsfig}
\usepackage{longtable}
\usepackage{supertabular}
\usepackage{color}
\usepackage{amssymb}

\usepackage{natbib}

 \usepackage{hyperref} 
\begin{document} 

\newcommand{\ms}{m\,s$^{-1}$\/}
\newcommand{\kms}{km\,s$^{-1}$\/}
\newcommand{\cmss}{cm\,s$^{-2}$\/}
\newcommand{\gcm}{g\,cm$^{-3}$\/}

\title{The GAPS Programme with HARPS-N at TNG \thanks{Based on observations collected at the Italian {\em Telescopio Nazionale Galileo} (TNG), 
operated on the island of La Palma by the Fundaci\'on Galileo Galilei of the Istituto Nazionale di Astrofisica (INAF) at the Spanish 
Observatorio del Roque de los Muchachos of the Instituto de Astrof\'isica de Canarias, in the frame of the programme 
{\em Global Architecture of Planetary Systems} (GAPS).}}
\subtitle{III: The retrograde orbit of HAT-P-18b} 
\author{M. Esposito  \inst{1,2}
\and E. Covino \inst{3}
\and L. Mancini \inst{4}
\and A. Harutyunyan \inst{5}
\and J. Southworth \inst{6}
\and K. Biazzo \inst{7}
\and D. Gandolfi \inst{7}
\and A. F. Lanza \inst{7}
\and M. Barbieri \inst{8}
\and A. S. Bonomo \inst{9}
\and F. Borsa \inst{10}
\and R. Claudi \inst{11}
\and R. Cosentino \inst{5}
\and S. Desidera \inst{11}
\and R. Gratton \inst{11}
\and I. Pagano \inst{7}
\and A. Sozzetti \inst{9}
\and C. Boccato \inst{11}
\and A. Maggio \inst{12}
\and G. Micela \inst{12}
\and E. Molinari \inst{5,13}
\and V. Nascimbeni \inst{11}
\and G. Piotto \inst{8,11}
\and E. Poretti \inst{10}
\and R. Smareglia \inst{14}
}
\institute{Instituto de Astrof\'isica de Canarias, C/ V\'ia L\'actea, s/n, E38205 - La Laguna (Tenerife), Spain  \email{mesposito@iac.es}  
\and Dep. de Astrof\'isica, Universidad de La Laguna, Avda. Astrof\'isico Francisco  S\'anchez, s/n 38206 La Laguna, Tenerife, Spain 
\and INAF -- Osservatorio Astronomico di Capodimonte, via Moiariello, 16, 80131 Naples, Italy  
\and Max-Planck-Institut f\"ur Astronomie, K\"onigstuhl 17, D-69117 Heidelberg, Germany 
\and Fundaci\'on Galileo Galilei - INAF, Rambla Jos\'e Ana Fernandez P\'erez, 7 38712 Bre\~na Baja, TF - Spain
\and Astrophysics Group, Keele University, Staffordshire,ST5 5BG, UK 
\and INAF -- Osservatorio Astrofisico di Catania, via S. Sofia 78, 95123 Catania, Italy  
\and Dipartimento di Fisica e Astronomia Galileo Galilei, Universit\`a di Padova, Vicolo dell'Osservatorio 2, I-35122, Padova, Italy 
\and INAF -- Osservatorio Astrofisico di Torino, Via Osservatorio 20, I-10025, Pino Torinese, Italy 
\and INAF -- Osservatorio Astronomico di Brera, Via E. Bianchi 46, 23807, Merate (LC), Italy 
\and INAF -- Osservatorio Astronomico di Padova, Vicolo dell'Osservatorio 5,  35122 Padova, Italy 
\and INAF -- Osservatorio Astronomico di Palermo, Piazza del Parlamento, Italy 1, I-90134, Palermo, Italy 
\and  INAF -- IASF Milano, via Bassini 15, I-20133 Milano, Italy 
\and INAF -- Osservatorio Astronomico di Trieste, Via Tiepolo 11,  34143 Trieste Italy 
}

\date{Received date /
Accepted date }

\abstract 
{The measurement of the Rossiter-McLaughlin effect for transiting exoplanets 
places constraints on the orientation of the orbital axis with respect to the stellar spin axis, 
which can shed light on the mechanisms shaping the orbital configuration of planetary
systems. 
 Here we present the interesting case of the Saturn-mass planet HAT-P-18b,
which orbits one of the coolest stars for which the Rossiter-McLaughlin effect
has been measured so far. 
 We acquired a spectroscopic time-series, spanning a full transit, with the HARPS-N spectrograph
mounted at the TNG telescope. The very precise radial velocity measurements delivered by the HARPS-N pipeline were used to
measure the Rossiter-McLaughlin effect. Complementary new photometric observations of another full transit were
also analysed to obtain an independent determination of the star and planet parameters. 
 We find that HAT-P-18b lies on a counter-rotating orbit, the sky-projected angle between the stellar spin axis and the planet
orbital axis being $\lambda=132\pm15$ deg. By joint modelling of the radial velocity and photometric data we obtain new
determinations of the star ($M_\star=0.770 \pm 0.027$ M$_\odot$;
 $R_\star=0.717 \pm 0.026$ R$_\odot$;  $V\sin{I_\star}=1.58  \pm 0.18$ \kms)
 and planet ($M_{\rm p}=0.196 \pm 0.008$ M$_{\rm J}$; $R_{\rm p}=0.947 \pm 0.044$ R$_{\rm J}$)
 parameters. Our spectra provide for the host star an effective temperature $T_{\rm eff}=4870 \pm 50$ K,
 a surface gravity of $\log g_\star=4.57 \pm 0.07 $ \cmss, and an iron abundance of 
 [Fe/H] = $ 0.10 \pm 0.06$. 
 HAT-P-18b is one of the few planets known to transit a star with $T_{\rm eff}\lesssim6250$ K on a retrograde orbit. Objects such as
HAT-P-18b (low planet mass and/or relatively long orbital period) most likely have a weak tidal coupling with their parent stars, therefore
their orbits preserve any original misalignment. As such, they are ideal targets to study the causes of orbital 
evolution in cool main-sequence stars.}


\titlerunning{GAPS. III. The retrograde orbit of HAT-P-18b}
\authorrunning{Esposito et al.}

\maketitle

\section{Introduction} \label{Sec:intro}

The number of known extrasolar planets has recently passed the milestone of one thousand.
While many discovery surveys are still ongoing, 
the characterization of known extrasolar
planetary systems is gaining ever more attention.
Transiting extrasolar planets (TEPs) are 
especially interesting as they allow for the
direct determination of fundamental parameters
such as planetary mass and radius  \citep{2012MNRAS.426.1291S}. Moreover, 
observations of secondary eclipses put constraints
on the planet albedo and brightness temperature,
while transmission spectroscopy can be used to probe
molecular and atomic features in the planet atmospheres.

Another possibility offered by TEPs
is to study the Rossiter-McLaughlin (RM) effect,
which is an anomaly in the radial velocity orbital trend
that occurs when the planet moves across the stellar photospheric disc
(see \citealt{2011ApJ...742...69H} and references therein).
The measurement of the RM effect permits the determination of the angle $\lambda$, 
the projection on the sky plane of the  misalignment angle $\Theta$ between the stellar spin axis 
and the planet orbital axis. The knowledge of $\lambda$ can give insight
into the mechanisms of formation and orbital migration of
exoplanets (\citealt{2011Natur.473..187N}; \citealt{2008ApJ...678..498N}; \citealt{2011ApJ...735..109W}).

In the context of GAPS,
a long-term observational programme with HARPS-N at TNG 
(\citealt{2013A&A...554A..28C}, hereafter Paper I; \citealt{2013A&A...554A..29D}),
we are carrying out a sub-programme aimed at measuring the 
RM effect in a sample
of TEP host stars.
We plan to explore a wide assortment of stellar temperatures, ages, and masses,
as well as diverse orbital (period, eccentricity) and physical
(mass, radius) planet properties.

In this paper, we report on the measurement of the RM effect
for the \object{HAT-P-18} transiting system \citep{2011ApJ...726...52H}. 
\object{HAT-P-18b}  is a Saturn-mass planet orbiting a K2 dwarf star 
with a period $P\sim5.5$ days.
\citet{2011ApJ...726...52H} (hereafter H11) pointed out that with a density $\rho_{\rm p}\sim0.25$ g cm$^{-3}$,
HAT-P-18b is not expected to
have a significant heavy element core, according to the planetary 
models by \citet{2007ApJ...659.1661F}.

\section{Observations and data reduction} \label{Sec:obs}

A time series of 20 spectra of HAT-P-18 was acquired with HARPS-N at TNG
\citep{2012SPIE.8446E..1VC} in 2013 June 11-12,  
bracketing a full transit of HAT-P-18b.
The exposure time was of 15 minutes, 
resulting in spectra with an S/N$\sim$20 (per extracted pixel at 5500 \AA),
degrading to $\sim$15 for the last three spectra due to
worsening seeing conditions (see Table\,\ref{t:RV-data}).
In the following months, three additional spectra were taken,
two of which at phases nearly corresponding
to the expected minimum and maximum of the radial velocity curve.

HARPS-N spectra were reduced using the standard pipeline. 
Radial velocities (RVs) were derived using the weighted cross-correlation function (CCF) method
(\citealt{1996A&AS..119..373B}, \citealt{2002A&A...388..632P}); for HAT-P-18
we used the K5 mask.
The pipeline also provided rebinned 1D spectra
that were used for the stellar atmosphere characterization (see Sec. \ref{SubSec:atmo}).

 A complete transit event of HAT-P-18b was observed on 2011 May 24,
with the Bologna Faint Object Spectrograph \& Camera (BFOSC) imager
mounted on the 1.52-m Cassini Telescope at the Astronomical
Observatory of Bologna in Loiano, Italy \citep{2013A&A...551A..11M,2013A&A...557A..30C}.
The night was not photometric and some data before ingress were
 rejected as they were affected by clouds. 
 The  CCD was used unbinned, giving a plate scale of
 $0.58^{\prime\prime}/\rm{pixel}$  for a total field-of-view of
 $13^{\prime} \times 12.6^{\prime}$, and the transit was observed
 through a Gunn $r$ filter. 
The telescope was autoguided and defocussed 
 to increase the exposure time to 140 sec, which minimisies the effects of
 systematic noises. 
The photometric data 
were derived using an upgraded version of the {\sc
 defot} package \citep{2009MNRAS.399..287S,2009MNRAS.396.1023S}.

\onltab{
\begin{table}
\caption{
HARPS-N RV measurements of HAT-P-18.
}
\label{t:RV-data}
\centering
\begin{tabular}{lcrrc}
\hline
\hline

 BJD (TDB)      &   RV      & error &   S/N\tablefootmark{a} &  \tablefootmark{b} \\ 
             & [\ms]    & [\ms] &                 &           \\
\hline
         2456455.451721   & $-$11094.3       &    5.2 &       20.7  &  o  \\
         2456455.462430   & $-$11099.6       &    4.8 &       21.9  &  o  \\
         2456455.473147   & $-$11094.8       &    6.3 &       17.8  &  o  \\
         2456455.483869   & $-$11093.4       &    7.0 &       16.6  &  o  \\
         2456455.494596   & $-$11095.1       &    5.9 &       18.7  &  o  \\
         2456455.505314   & $-$11106.0       &    4.7 &       21.7  &  i  \\
         2456455.516031   & $-$11115.9       &    4.5 &       22.5  &  i  \\
         2456455.526749   & $-$11116.5       &    4.9 &       20.8  &  i  \\
         2456455.537462   & $-$11103.8       &    4.5 &       22.4  &  i  \\
         2456455.548184   & $-$11118.4       &    5.2 &       20.2  &  i  \\
         2456455.558915   & $-$11091.6       &    7.2 &       16.1  &  i  \\
         2456455.569633   & $-$11100.7       &    5.2 &       20.3  &  i  \\
         2456455.580346   & $-$11095.8       &    4.4 &       23.0  &  i  \\
         2456455.591059   & $-$11096.2       &    4.7 &       22.2  &  i  \\
         2456455.601781   & $-$11088.2       &    4.7 &       22.0  &  i  \\
         2456455.612495   & $-$11101.9       &    4.6 &       22.6  &  i  \\
         2456455.623212   & $-$11102.1       &    5.1 &       20.8  &  o  \\
         2456455.633939   & $-$11095.6       &    7.4 &       15.8  &  o  \\
         2456455.644665   & $-$11122.4       &    8.3 &       15.1  &  o  \\
         2456455.656845   & $-$11108.6       &    7.1 &       16.2  &  o  \\
         2456506.466435   & $-$11130.3       &    6.8 &       17.3  &  o  \\
         2456536.490373   & $-$11073.4       &    5.8 &       19.5  &  o  \\
         2456543.375155   & $-$11095.1       &    4.5 &       23.7  &  o  \\
\hline 
\end{tabular}
\tablefoot{ 
\tablefoottext{a}{per pixel at 5500 \AA}
\tablefoottext{b}{i\,$\equiv$\,in-transit, o\,$\equiv$\,out-of-transit}
}
\end{table}
}

\section{Results} \label{Sec:res}

\subsection{Spectroscopic determination of stellar parameters} \label{SubSec:atmo}

 We derived the photospheric parameters of the planet-hosting star HAT-P-18 by applying two different methods 
 on the mean of all the available HARPS-N spectra. 

 The first method relies on the use of the spectral analysis package MOOG (\citealt{1973ApJ...184..839S}, version 2013). 
 As in Paper I, we measured the equivalent widths (EWs) of iron lines chosen from the list 
 by \citet{2012MNRAS.427.2905B} and adopted the {\it abfind} driver within MOOG. We hence determined  the effective 
 temperature ($T_{\rm eff}$) by imposing that the \ion{Fe}{i}  abundance does not depend on the excitation 
 potential of the lines, the microturbulence velocity ($v_{\rm mic}$) by imposing that the \ion{Fe}{i} abundance 
 is independent on the EW of the lines, and the surface gravity ($\log\,g_*$) by the \ion{Fe}{i}/\ion{Fe}{ii} ionization 
 equilibrium condition. The projected rotational velocity $V$\,sin\,$I_{\star}$ was measured following the procedure 
 described in \citet{2011A&A...526A.103D}.

 The second method compares the composite HARPS-N spectrum with a grid of theoretical model spectra 
  using spectral features that are sensitive to different photospheric parameters 
 (\citealt{2004astro.ph..5087C}; \citealt{2005A&A...443..735C}; \citealt{2008A&A...486..951G}).
 Briefly, we used the wings of the Balmer lines to estimate the $T_{\rm eff}$ of the star, 
 and the Mg\,{\sc i} 5167, 5173, and 5184~\AA, the Ca\,{\sc i} 6162 and 6439~\AA, and the Na\,{\sc i} D lines to determine 
 its $\log\,g_*$. The iron abundance [Fe/H] and $v_{\rm mic}$ were derived by applying the method described 
 in \citet{1979MNRAS.186..673B} on isolated \ion{Fe}{i} and \ion{Fe}{ii} lines. The $V$\,sin\,$I_{\star}$ and macroturbulence 
 velocity ($v_{\rm mac}$) were measured by fitting the profiles of several clean and isolated metal lines.
 
 The two methods provided consistent results, well within the error bars. The final adopted values,
 obtained as the weighted mean of the two independent determinations, 
 agree very well with the values by H11
 (see Table \ref{table_par}).
  We note that the $V$\,sin\,$I_{\star}$ and $log\,g_*$ are consistent within the errors with the values 
 obtained by modelling the RM- and light-curve (see Sec. \ref{SubSec:fit}), thereby validating our global analysis.

\subsection{RV and photometric data analysis} \label{SubSec:fit}

The RV and photometric data sets were analysed jointly. To this
purpose we developed a MATLAB$^{\textregistered}$ code that implements a global model 
and a data-fitting algorithm.

The model considers the parameters necessary
to describe the planet and star position and velocity vectors
at any given time, that is, the masses of the star $M_\star$ and of the planet $M_{\rm p}$,
the orbital period $P$ and eccentricity $e$, the epoch $\tau$ and argument $\omega$ of
periastron, the systemic RV $\gamma$; the orbital space orientation is
described by the inclination angle $i_p$ and the misalignment angle $\lambda$
\footnote{The third angle, the longitude of the ascending node, is not considered as it does
not affect the RV and photometric measurements; in other words, it is not an observable.}.
Other parameters necessary to model the RM effect and the light curve are the stellar $R_\star$
and  planetary $R_{\rm p}$ radius, the stellar projected rotational velocity $V\sin{I_\star}$,
and the limb-darkening coefficients. Our model can implement each of the five equations
proposed by \citet{2011A&A...529A..75C} to describe the limb-darkening law.
Other effects that can affect the measurements, such as 
stellar surface inhomogeneities, stellar differential
rotation and convective blue-shift, are not included in the model.
We refer to Paper I for the details of the method
used to determine the RV anomaly when the planet is transiting the
stellar disc.

The best-fit values of the parameters are obtained by a
least-squares minimization algorithm. The region of the
parameters space to be explored can be limited
providing upper and lower limits to the parameter values.
Most importantly, any number of linear and non-linear constraints
can be set: this allows placing limits on other parameters
(such as $K$, $T_{14}$, $b$, see Table \ref{table_par} for their definition), even though they are not direct parameters
of the fit.
The mass of the star, ($M_\star =  0.770 \pm 0.027 M_\odot$), is preliminarily determined from 
evolutionary track models \citep{2001ApJS..136..417Y}, adopting the values of the atmospheric parameters
determined previously and using the $a/R_\star$ value derived from the light curve. Evolutionary tracks
also provide an estimate of the stellar age of 7.0 $\pm$ 3.6 Gyr.

\begin{figure}
  \resizebox{\hsize}{!}{\includegraphics{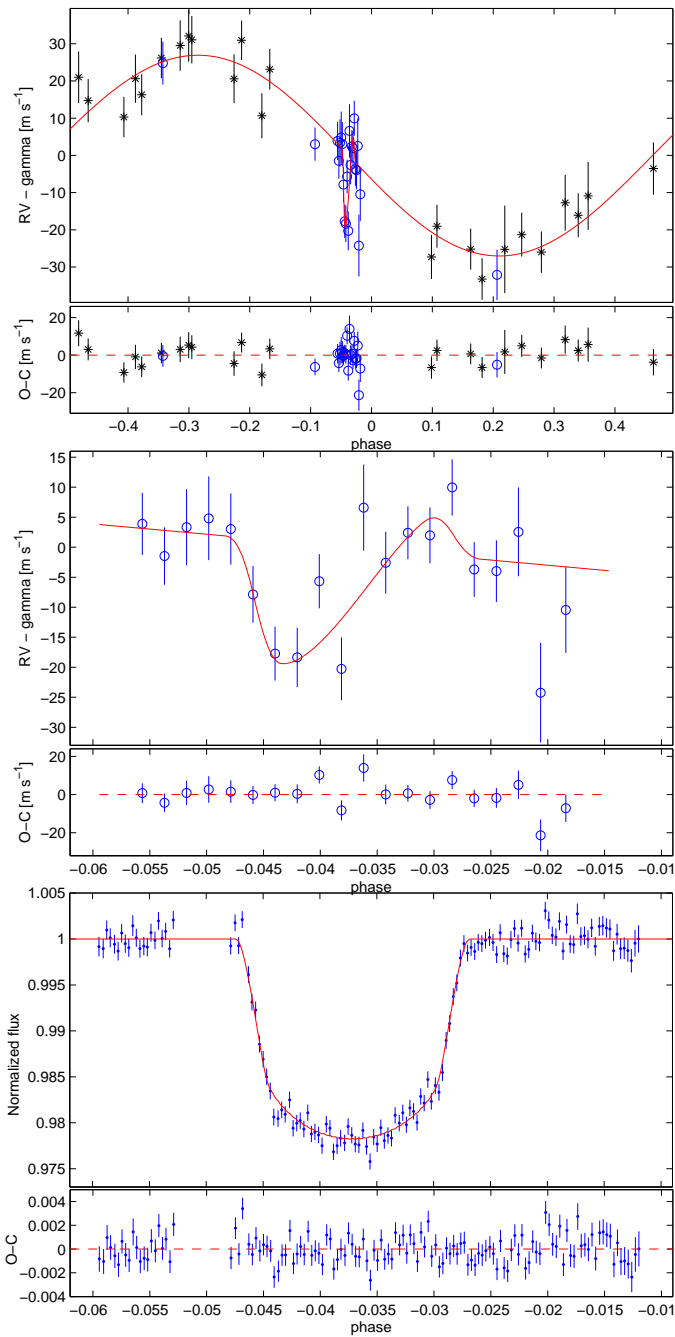}}
  \caption{{\it Upper panel}: phase-folded RV data set. 
  Blue open circles are the HARPS-N data, black asterisks are the RVs
  from Hartman et al. (2011). Superimposed is the best-fit RV curve. 
  {\it Middle panel}: zoom in the RV measurements spanning the transit. 
  {\it Lower panel}: $r$-band photometric time-series and best-fit light curve.}
  \label{Fig1}
\end{figure}

Together with our data sets, the global fit also considers the RVs presented in H11.
We show in the top panel of Fig. \ref{Fig1} the phase-folded RV data,
with superimposed the best-fit RV curve. We find the eccentricity to be
$ 0.009^{+0.03}_{-0.009}$, consistent with a circular orbit \citep{2011MNRAS.410.1895Z}.
The middle panel displays an expanded view of the phases around the transit. 
During the transit the RVs
are first blue- and then red-shifted with respect to the orbital trend, 
indicating that the planet is moving on a retrograde orbit. The best-fit value
for the sky-projected spin-orbit misalignment angle is $\lambda=132\pm15$ deg.
 To evaluate the significance of the detection of the RM effect, we
used the transit RV data alone (-0.050$<$phase$<$-0.024)
and compared the $\overline{\chi}^2$ values obtained by modelling the effect, $\overline{\chi}^2$=1.08,
and by just fitting the orbital trend, $\overline{\chi}^2$=4.43. 
The bottom panel of Fig. \ref{Fig1} shows the phase-folded $r$-band photometric data
set and the best-fit light curve. 
We adopted the simple linear law to describe the stellar limb-darkening,
as no significant improvement on the light curves fit is obtained by
using a quadratic law.

The best-fit values for all the parameters are listed in Table \ref{table_par},
together with the errors that were determined by means of a Monte Carlo method.
Our results agree well with those reported in H11.

\begin{table}
\caption{Star and planet parameters of the system HAT-P-18}              
\label{table_par}      
\centering                                      
\begin{tabular}{l c}          
\hline\hline                        
Parameter [Units]                                 &     Value   \\
\hline
\multicolumn{2}{c}{ {\it Stellar spectra characterization} } \\
\hline
Effective temperature, $T_{\rm eff}$ [K]                    &     $ 4870 \pm 50$      \\
Surface gravity, $\log g_\star$ [\cmss]                     &     $ 4.57 \pm 0.07$    \\
Iron abundance, [Fe/H]                                         &     $ 0.10 \pm 0.06$    \\
Microturbulence, $v_{\rm mic}$ [\kms]                       &     $ 0.89 \pm 0.08$    \\
Macroturbulence, $v_{\rm mac}$ [\kms]                       &     $ 2.5 \pm 0.5$    \\
Proj. rot. vel., $V\sin{I_\star}$ [\kms  ]                  &     $1.40   \pm 0.35$     \\ 
\multicolumn{2}{c}{ {\it RV and photometric data fit} } \\
\hline
Star mass, $M_\star$ [M$_\odot$]                            &    $ 0.770 \pm 0.027$   \\ 
Planet mass, $M_{\rm p}$ [M$_{\rm J}$]                      &    $ 0.196 \pm 0.008$   \\    
Star radius, $R_\star$ [R$_\odot$]                          &    $ 0.717 \pm 0.026$   \\ 
Planet radius, $R_{\rm p}$ [R$_{\rm J}$]                    &    $ 0.947 \pm 0.044$   \\ 
Orbital period, $P$ [days]                                  &    $ 5.507978 \pm 0.000043$ \\
Eccentricity, $e$                                           &    $ 0.009^{+0.03}_{-0.009}$      \\
Longitude of periastron, $\omega$ [deg]                     &    $ 104 \pm 50$      \\
Orbital inclination, $i_{\rm p}$ [deg]                            &    $ 88.79 \pm 0.25$  \\ 
Epoch of periastron, $\tau$ [BJD]                           &    $ 2455706.7 \pm 0.7$ \\
Barycentric RV, $\gamma$ [\ms]                              &    $ -11098.2 \pm 2.5$  \\
H11 RVs offset, $\gamma_2$ [\ms]                                  &    $-0.3 \pm 1.0$  \\
Proj. spin-orbit angle, $\lambda$ [deg]                     &    $ 132 \pm 15$         \\
Proj. rot. vel., $V\sin{I_\star}$ [\kms]                    &    $ 1.58  \pm 0.18$     \\ 
Limb dark. --- HARPS-N band, $u_{\rm RV}$                       &    $ 0.58 \pm 0.12$       \\
Limb dark. --- $r$ band, $u_{\rm r}$                           &    $ 0.56 \pm 0.07$      \\
Normalized chi-square, $\overline{\chi}^2$                              &   1.39     \\
\multicolumn{2}{c}{ {\it Derived parameters} } \\
\hline
Orbital semi-major axis, $a$ [AU]                           &    $ 0.0559\pm0.0007$    \\  
Transit duration, $T_{14}$ [hours]                          &    $ 2.69 \pm 0.06$      \\   
Impact parameter, $b$                                       &    $ 0.352 \pm 0.057$    \\
Transit depth --- $r$ band,                                   &    $ 0.02179 \pm 0.00048$  \\  
RV-curve semi-amplitude, $K$ [\ms]                          &    $ 26.9 \pm 1.1$     \\
Star density, $\rho_\star$ [\gcm]                           &    $ 2.94 \pm 0.30$    \\
Star surface gravity, $\log g_\star$ [\cmss]                &    $ 4.613 \pm 0.031$   \\
Planet density, $\rho_{\rm p}$ [\gcm]                             &    $ 0.286 \pm 0.042$   \\
Planet surface gravity, $\log g_{\rm p}$ [\cmss]                  &    $ 2.734  \pm 0.044$    \\
Planet equilibrium temperature, $T_{\rm p}$ [K]             &    $ 841 \pm 15 $        \\  
\hline                                             

\end{tabular}
\end{table}

\section{Discussion} \label{Sec:disc}

\begin{figure}
  \resizebox{\hsize}{!}{\includegraphics{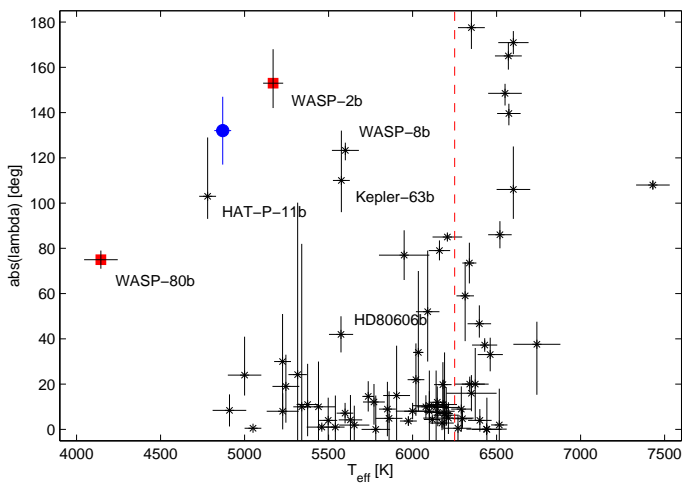}}
  \caption{Compilation of the values of $\lambda$, measured via the RM effect,
  as a function of the host star effective temperature 
  (see: http://www.astro.keele.ac.uk/jkt/tepcat/rossiter.html).
  HAT-P-18b is shown as a filled blue dot.
 For the two objects marked with red squares, the determination of $\lambda$
  is doubtful (see text for details).
  The red dashed line at $T_{\rm eff}$ = 6250 K marks the limit above which the
  mass of the convective shell becomes negligible \citep{2010ApJ...718L.145W}.}
  \label{Fig2}
\end{figure}

\citet{2010ApJ...718L.145W} first noticed, and  later \citet{2012ApJ...757...18A} (hereafter A12)
confirmed, an empirical correlation
between the spin-orbit relative orientation and the effective temperature
of the host star:
planets hosted by stars with $T_{\rm eff}$ $\gtrsim$ 6250 K display
a wide distribution of $\lambda$ values,
while planets around cooler stars are almost always well aligned (see Fig. \ref{Fig2}).

This fact has been interpreted as supporting evidence of a scenario in which giant planets 
approach closely their parent stars following planet-planet gravitational scattering, 
Kozai-Lidov cycles, or secular chaotic orbital evolution, 
as opposed to gentle migration in a protoplanetary disc 
(\citealt{2013ApJ...767L..24D}; \citealt{2011ApJ...735..109W}; \citealt{2013arXiv1312.4293B}). 
Initially, planets can have large misalignments around both cool and hot stars. 
Later on, because of their convective envelopes, tidal interactions 
are effective in cool stars to realign the systems on relatively short time-scales.

 HAT-P-18, with a  $T_{\rm eff}$=$ 4870 \pm 50$ K
and a very large misalignment angle $\lambda=132\pm15$ deg,
seems to represent an exception in this scheme.
However, as already argued by A12,
the realignment time-scale also depends on other
parameters such as the planet/star mass ratio and the
scaled semi-major axis $a/R_\star$. In fact, using the equation (2) in 
A12 to estimate the alignment timescale $\tau_{\rm CE}$,
for HAT-P-18 we find $\tau_{\rm CE} \backsimeq  10^{15}$ yr, adopting the equilibrium tide theory of \citet{1977A&A....57..383Z}. 
Given the present uncertainty of the tidal theory, rather than the absolute value of $\tau_{\rm CE}$, 
it is interesting how it compares with the values for other TEPs.
HAT-P-18b has one of the highest values of $\tau_{\rm CE}$ among the TEPs
that orbit cool stars, inferior only to \object{HAT-P-11b} and \object{HD80606b}, which 
are also both significantly misaligned (see Fig. 24 in A12).
Therefore, HAT-P-18b confirms that  planets around cool stars
can also have misaligned orbits, provided that 
they have a longer tidal realignment time-scale due
to smaller mass 
 (HAT-P-11b, \citealt{2010ApJ...723L.223W}; \object{Kepler-63b}, \citealt{2013ApJ...775...54S}) and/or larger orbital
semi-major axis (\object{WASP-8b}, \citealt{2010A&A...517L...1Q}; HD80606b, \citealt{2010A&A...516A..95H};  Kepler-63b). 
The high misalignment angle $\lambda=153$ deg of \object{WASP-2b} \citep{2010A&A...524A..25T} is at odds
with its mass and separation ($M_{\rm p}$=0.9 M$_{\rm J}$, $a$=0.03 AU),
but the reliability of the measure was confuted by \citet{2011ApJ...738...50A},
based on new observations.
The value of $\lambda$ for \object{WASP-80b} \citep{2013A&A...551A..80T} is strongly dependent on the value assumed
for $V\sin{I_\star}$, because of the nearly zero impact parameter. 

The circularization of the orbit proceeds on a much shorter time-scale because 
of tidal dissipation inside the planet. Adopting a modified tidal quality factor of $Q^{\prime}_{\rm p} = 10^{5}$, 
corresponding to the value measured in Jupiter \citep{2009Natur.459..957L}, we obtain a damping time-scale of about 90 Myr 
for the eccentricity. 

\citet{2013ApJ...769L..10R} challenged the  
interpretation of A12 of the $\lambda$--$T_{\rm eff}$ correlation.
They advocated migration in the protoplanetary disc 
that produces aligned hot-Jupiters,
and invoked a mechanism based on stellar internal gravity waves
to explain the high obliquities found in hot stars.
We point out that this mechanism is not applicable 
to stars such as HAT-P-18, which have a convective envelope.

\section{Conclusions} \label{Sec:concl}
We have found that the Saturn-mass planet hosted by  HAT-P-18,
a K2 dwarf star with $T_{\rm eff} =  4870 \pm 50$ K,                       
lies on a retrograde orbit. We discussed how the
existence of such object fits in the context of 
the current alternative theories of giant planet orbital
migration. 
HAT-P-18b scores a point in favour of gravitational
N-body (N$\geqslant$3) interactions, while migration in the proto-planetary disc
seems unable to explain its existence.
HAT-P-18b, which is one of the very few planets around cool stars found 
to be on a retrograde orbit, also allows setting constraints
on the efficiency of tidal interactions in obliquity damping. 

\begin{acknowledgements} 
ME acknowledges financial support from the Spanish Ministry
project MINECO AYA2011-26244.
We thank the TNG staff for help in 
the observations and 
with data retrieval from the TNG archive. 
The GAPS project in Italy acknowledges the support by INAF through 
the "Progetti Premiali" funding scheme of the Italian Ministry of Education, University, and Research.
DG acknowledges funding from the 
European Union Seventh Framework Programme (FP7/2007-2013) under grant agreement n. 267251.
\end{acknowledgements}

\vspace{-0.6cm}

\bibliographystyle{aa} 
\bibliography{list_ref_hatp18.bib} 

\end{document}